\documentclass{PoS}

\usepackage{lineno}

\title{Status, open problems and prospects of the decay $B^+\rightarrow \ell^+ \nu_{\ell}$}

\ShortTitle{HQL 2010}

\author{
\speaker{Alejandro P\'erez}\\
On behalf of the SuperB collaboration \\
INFN Sezione di Pisa\\
        E-mail: \email{luis.alejandro.perez@pi.infn.it}}

\abstract{The study of rare B-decays at SuperB provides unique opportunities to understand the 
Standard Model (SM) and constrain new physics (NP). We discuss the new physics potential of the 
leptonic $B^+ \rightarrow \ell^+\nu_{\ell}$ decays from the proposed SuperB experiment with $75{\rm ab^{-1}}$ 
of data (5 nominal years of data taking).}

\FullConference{The Xth Nicola Cabibbo International Conference on Heavy Quarks and Leptons,\\
		October 11-15, 2010\\
		Frascati (Rome) Italy}

\begin{document}


\section{Introduction}

SuperB~\cite{SuperB} is a high luminosity $e^+e^-$ collider that will be able to indirectly probe NP 
at energy scales far beyond the reach of any accelerator planned or in existence. Just as detailed 
understanding of the SM was developed from stringent constraints imposed by flavour changing processes between 
quarks, the structure of any NP is severely constrained by flavour processes. The pattern of deviations 
from the SM can be used to test the NP. If NP is found at the LHC, then the many 
golden measurements from SuperB (of which $B^+ \rightarrow \ell^+\nu_{\ell}$ is an example) will help decode 
the subtle nature of the NP. However if no new particles are found at the LHC, SuperB will be able to search 
for NP at energy scales up to $100~{\rm TeV}$. In either scenario, flavour physics measurements that can be 
made at SuperB play an important role in understanding the nature of NP.

In the SM the purely leptonic $B$ meson decays $B^{+} \rightarrow \ell^{+}\nu_{\ell}$ proceed 
at the lowest order through an annihilation diagram with a $W^{+}$ exchange. The SM branching ratio (${\rm BR}$) 
can be calculated as~\cite{Silverman}
\begin{equation}
\label{eq:BF_Blnu}
{\rm BR}(B^{+}\rightarrow \ell^{+}\nu_{\ell})_{\rm SM} = \frac{G^2_F m_B m^2_{\ell}}{8\pi}\left(1 - \frac{m^2_{\ell}}
{m^2_B} \right)^2 f^2_B |V_{ub}|^2\tau_B~,
\end{equation}
where $G_F$ is the Fermi constant, $m_{\ell}$ and $m_B$ are the lepton and $B^+$ masses, respectively, 
and $\tau_B$ is the $B^+$ lifetime. The ${\rm BR}$ is sensitive to the CKM matrix element $|V_{ub}|$~\cite{CKMmatrix}
and the $B$ decay constant $f_B$.

The SM estimate for ${\rm BR}(B^{+}\rightarrow \tau^{+}\nu_{\tau})$ is $(1.20 \pm 0.25)\times10^{-4}$, this 
assuming $\tau_B = 1.638\pm0.011~{\rm ps}$~\cite{TauB}, $|V_{ub}| = (4.32 \pm 0.16 \pm 0.29)\times10^{-3}$ 
(errors are statistical and systematic, respectively)~\cite{HFAG}, and $f_B = 190 \pm 13 {\rm MeV}$~\cite{HPQCD}. 
The main uncertainties on the expected SM ${\rm BR}$ come from the $|V_{ub}|$ and $f_B$ parameters. 
To a very good approximation, helicity is conserved in $B^{+} \rightarrow \mu^+\nu_{\mu}$ and $B^{+} \rightarrow e^+\nu_{e}$ 
decays, leading to ${\rm BR}(B^{+}\rightarrow \mu^{+}\nu_{\mu}) = (5.4\pm1.1)\times10^{-7}$ and 
${\rm BR}(B^{+}\rightarrow \mu^{+}\nu_{\mu}) = (1.3\pm0.4)\times10^{-11}$. However, reconstruction of 
$B^+\rightarrow \tau^+\nu_{\tau}$ decays is experimentally more challenging than $B^+\rightarrow \mu^+\nu_{\mu}$ or 
$B^+\rightarrow e^+\nu_{e}$ due to the large missing momentum from multiple neutrinos in the final state.

Purely leptonic $B$ decays are sensitive to NP, where additional heavy virtual particles replace the 
$W^{+}$ and contribute to the annihilation processes. Charged Higgs boson effects may 
greatly enhance or suppress the decay rate in some two-Higgs-doublet models~\cite{TwoHiggsDoubletModel}.
Similarly, there may be enhancements through mediation of leptoquarks in the Pati-Salam model of quark-lepton 
unification~\cite{SUSY}. Direct test of Yukawa interactions in and beyond the SM are possible in the study of 
these decays, as annihilation processes proceed through the longitudinal component of the intermediate vector 
boson. In particular, in a SUSY scenario at large $\tan\beta$, non-SM effects in helicity-suppressed charged 
current interactions are potentially observable, being strongly $\tan\beta$-dependent and leading 
to~\cite{TwoHiggsDoubletModel}
\begin{equation}
\label{eq:BF_Blnu_NP}
\frac{{\rm BR}(B^+\rightarrow \ell^+\nu_{\ell})_{\rm NP}}{{\rm BR}(B^+\rightarrow \ell^+\nu_{\ell})_{\rm SM}} \simeq 
\left( 1 - \tan^2\beta\frac{m^2_B}{M^2_H} \right)^2~,
\end{equation}
where $M_H$ is the charged Higgs mass and ${\rm BR}(B^+\rightarrow \ell^+\nu_{\ell})_{\rm NP}$ is the NP expectation 
in the before mentionned NP models. As can be see from eq.~\ref{eq:BF_Blnu_NP}, a measurement of the ${\rm BR}$ allows 
to set a constraint on the $\tan\beta - M_H$ plane.

\section{Experimental Technique}

The recoil technique has been developed in order to search for rare $B$ decays with undetected particles, 
like neutrinos, in the final state. The technique consists of the reconstruction of one of the two B mesons 
($B_{\rm tag}$), produced through the $e^+e^- \rightarrow \Upsilon(4S) \rightarrow B\bar{B}$ resonance,
 in a high purity hadronic or semi-leptonic final states, allowing to select a pure sample of 
$B\bar{B}$ events. Having identified the $B_{\rm tag}$, everything in the rest of the event (ROE) belongs by 
default to the signal B candidate ($B_{\rm sig}$), and so this technique provides a clean environment to 
search for rare decays. In this analysis, the $B_{\rm tag}$ is reconstructed in the hadronic modes (HD) 
$B\rightarrow D^{(*)}X$, where $X = n\pi + mK + pK^0_S + q\pi^0$ ($n+m+p+q < 6$), or semi-leptonic modes (SL)
$B\rightarrow D^{(*)}\ell\nu$, ($\ell = e,~\mu$).

In the search for $B^+ \rightarrow \mu^+ \nu_{\mu}$ and $B^+ \rightarrow e^+ \nu_{e}$ decays, the signal 
is given by a single track identified as a muon and electron, respectively, in the ROE. In the search of 
$B^+ \rightarrow \tau^+ \nu_{\tau}$ decays, a single track as a muon, electron or pion is selected from the ROE, 
compatible with the $\tau^+ \rightarrow \mu^+ \nu_{\mu} \bar{\nu}_{\tau}$, $\tau^+ \rightarrow e^+ \nu_{e} \bar{\nu}_{\tau}$ 
and $\tau^+ \rightarrow \pi^+ \bar{\nu}_{\tau}$ decays, respectively. Furthermore, a single track and a neutral pion 
in the ROE is searched to reconstruct $\rho^+\rightarrow \pi^+\pi^0$ candidates compatible with 
the $\tau^+ \rightarrow \rho^+ \bar{\nu}_{\tau}$ decay.

One very important variable is the lepton momentum ($p'_{\ell}$) in the $B_{\rm sig}$ rest-frame, 
as the $B^+\rightarrow \ell^+\nu_{\ell}$ channels ($\ell = e,\mu$) produce monoenergetic leptons. 
This variable allows to separate $B^+\rightarrow \ell^+\nu_{\ell}$ from 
$B^+\rightarrow \tau^+(\rightarrow \ell^+ \nu_{\ell} \bar{\nu}_{\tau})\nu_{\tau}$ events, and provides 
additional discrimination against other sources of background. The closed kinematics of the hadronic 
recoil technique allow to easily calculate the $B_{\rm sig}$ rest frame from the reconstructed $B_{\rm tag}$ 
and beam energies. However, the semi-leptonic recoil technique poses a problem due to the presence of a 
neutrino in the $B_{\rm tag}$ reconstruction. As the only missing particle in the $B_{\rm tag}$ is a 
neutrino, it is possible to calculate CM angle between the $B_{\rm tag}$ and $D^{(*)}\ell$ momenta. Yet, 
as the $B_{\rm sig}$ and $B_{\rm tag}$ are back-to-back in the CM frame, this means that the $B_{\rm sig}$ 
momentum is contained in a cone around the $D^{(*)}\ell$ system. Using this information and the magnitude of 
the $B_{\rm sig}$ CM momentum ($p^*_{B} = \sqrt{(E^*_{\rm beam}/2)^2 - m^2_{B}}$, with $E^*_{\rm beam}$ the 
total beam energy in the CM-frame), it is possible to construct an estimator of $p'_{\ell}$ as the arithmetic 
average of the $p'_{\ell}$ calculated using all possible $B_{\rm sig}$ directions around the $D^{(*)}\ell$ 
system.

Finally, for these kind of decay modes with undetected particles in the final state, the most powerful 
variable for separating signal and background is the so-called extra energy, $E_{\rm extra}$, which is 
defined as the extra energy in the electromagnetic calorimeter not associated with the $B_{\rm tag}$ or 
$B_{\rm sig}$ candidates. For the signal this variable peaks strongly near zero.

\section{Current Experimental Status}

The latest state of the art results on $B^+ \rightarrow \ell^+\nu_{\ell}$ decay rates from both BaBar 
and Belle collaborations are summarized in table~\ref{tab:Expe_measurements}. The current best knowledge 
on $B^+ \rightarrow \mu^+\nu_{\mu}$ and $B^+ \rightarrow e^+\nu_{e}$ channels are upper limits at $90\%$~C.L. 
In contrast, the $B^+ \rightarrow \tau^+\nu_{\tau}$ channel is a well established decay, with a value 
$(1.64 \pm 0.34)\times10^{-4}$ (combining all the experimental findings~\cite{HFAG}), which is in agreement 
with the SM expectation. However, this last experimental result is a source of tension within the the CKM 
global fit. The indirect determination of $B^+ \rightarrow \tau^+\nu_{\tau}$ turns out to be at $~2.6\sigma$ 
($~3.2\sigma$) from the experimental value, as estimated by the CKMfitter~\cite{CKMFitter} 
(UTfit~\cite{UTFIT}) collaboration. More precise experimental findings are needed to disentangle the current 
state of affairs.

\begin{table}[t]
\begin{center}
\begin{TableSize}
\begin{tabular}{l|cc}
\hline
Observable         & BaBar & Belle \\
\hline
${\rm BR}(B^+\rightarrow \tau^+\nu_{\tau})~~{\rm (SL)}$ & $(1.7 \pm 0.8 \pm 0.2)\times 10^{-4}$~\cite{BaBar_BToenu_BTomunu_BTotaunu_SL} & $(1.54^{+0.38+0.29}_{-0.37-0.31}\times 10^{-4}$~\cite{Belle_BTotaunu_SL} \\
${\rm BR}(B^+\rightarrow \tau^+\nu_{\tau})~~{\rm (HD)}$ & $(1.8^{+0.57}_{-0.54} \pm 0.26)\times 10^{-4}$~\cite{BaBar_BTotaunu_HD} & $(1.79^{+0.56+0.46}_{-0.49-0.51})\times 10^{-4}$~\cite{Belle_BTotaunu_HD} \\

${\rm BR}(B^+\rightarrow e^+\nu_{e})~~{\rm (SL)}$       & $<0.8\times 10^{-5}$~\cite{BaBar_BToenu_BTomunu_BTotaunu_SL} & --- \\
${\rm BR}(B^+\rightarrow e^+\nu_{e})~~{\rm (HD)}$       & $<1.9\times 10^{-6}$~\cite{BaBar_BToenu_BTomunu_HD} & $< 0.98\times 10^{-6}$~\cite{Belle_BToenu_BTomunu_HD} \\

${\rm BR}(B^+\rightarrow \mu^+\nu_{\mu})~~{\rm (SL)}$   & $<1.1\times 10^{-5}$~\cite{BaBar_BToenu_BTomunu_BTotaunu_SL} & --- \\
${\rm BR}(B^+\rightarrow \mu^+\nu_{\mu})~~{\rm (HD)}$   & $<1.0\times 10^{-6}$~\cite{BaBar_BToenu_BTomunu_HD} & $< 1.70\times 10^{-6}$~\cite{Belle_BToenu_BTomunu_HD} \\
\hline
\end{tabular}
\caption{\em Summary of the experimental findings on $B^+ \rightarrow \ell^+\nu_{\ell}$. The first and second 
errors are statistical and systematic. Upper limits are at $90\%$ C.L.}
\label{tab:Expe_measurements}
\end{TableSize}
\end{center}
\end{table}

\section{SuperB detector layout studies}

Even though the expected SuperB increase in the instantaneous luminosity of a factor of $~100$ already promises 
significant improvements on the leptonic $B^+ \rightarrow \ell^+\nu_{\ell}$ decays, additional activities for 
detector optimization are currently ongoing. The SuperB baseline detector configuration is very similar to BaBar 
but the boost ($\beta\gamma$) is reduced from $0.56$ to $0.28$. This reduction increases the geometrical acceptance 
and so the reconstruction efficiency. A new layer is added to the vertex detector as close as possible to the beam
pipe in order to not to degrade the time-dependent measurements. Additionally, the inclusion of two new devices 
that will increase further the geometrical acceptance of the detector is beging considered: a particle identification 
device (Fwd-PID) placed in the fordward region and an electromagnetic calorimeter (Bwd-EMC) located in the backward region, 
covering the polar angular regions of $(17,25)$ and $(152,167)$ degrees, respectively.

The Fwd-PID is a highly performant PID device for $K/\pi$ separation based on time-of-flight measurements, 
located in a region previously covered only by the tracking system. This new device will improve particle 
identification in a momentum region from $1.6$ to $5.0~{\rm GeV}$ where the tracking system alone is 
poorly performant. The Bwd-EMC will be used as a veto device, which means that no neutrals measured in 
it will be used to reconstruct the $B_{\rm tag}$ and $B_{\rm sig}$ candidates. Additional background 
suppression can be achieved by cutting on the total energy deposited in the Bwd-EMC, as the signal is 
expected to peak strongly at zero.

The SuperB fast simulation has been used to produce signal and the main background (generic $B\bar{B}$ decays) 
samples in the previously mentioned detector setups. This test showed that the reduced boost has the effect of 
increasing the signal efficiency by $\sim 7\%$ with an additional background suppression of $\sim 6\%$. 
The impact of the Fwd-PID device is to increase the signal and background reconstruction efficiencies by the same 
amount of $~2.5\%$, due to an increase of the tag-side kaons identification efficiency in the forward region.  
Finally, the impact of the Bwd-EMC is to reduce the backgrounds by $\sim 10\%$ with a negligible 
effect on the signal. The total effect is, at a fixed integrated luminosity, an increase in the total sample 
efficiency with a higher signal to background ratio $S/B$.

\section{Expected sensitivities}

The $S/\sqrt{(S+B)}$ ratio, which would be the statistical significance of the ${\rm BR}$ measurement in a 
{\it cut-and-count} analysis, can be used as a measure of the expected sensitivities in SuperB. This ratio only 
takes into account the statistical uncertainties, and needs to be modified in order to consider the irreducible 
systematic uncertainties,
\begin{equation}
\label{eq:Significane_syst}
{\rm Significance} = \frac{S}{\sqrt{(S+B+(\epsilon_{\rm syst}S)^2)}}~,
\end{equation}
where $\epsilon_{\rm syst}$ is the total relative systematic error. No significant observation is expected at 
SuperB of the highly suppressed $B^+ \rightarrow e^+\nu_e$ decay, therefore it will be excluded from the 
subsequent discussion.

\begin{figure}[ht!]
\begin{center}
\includegraphics[width=7.0cm,keepaspectratio]{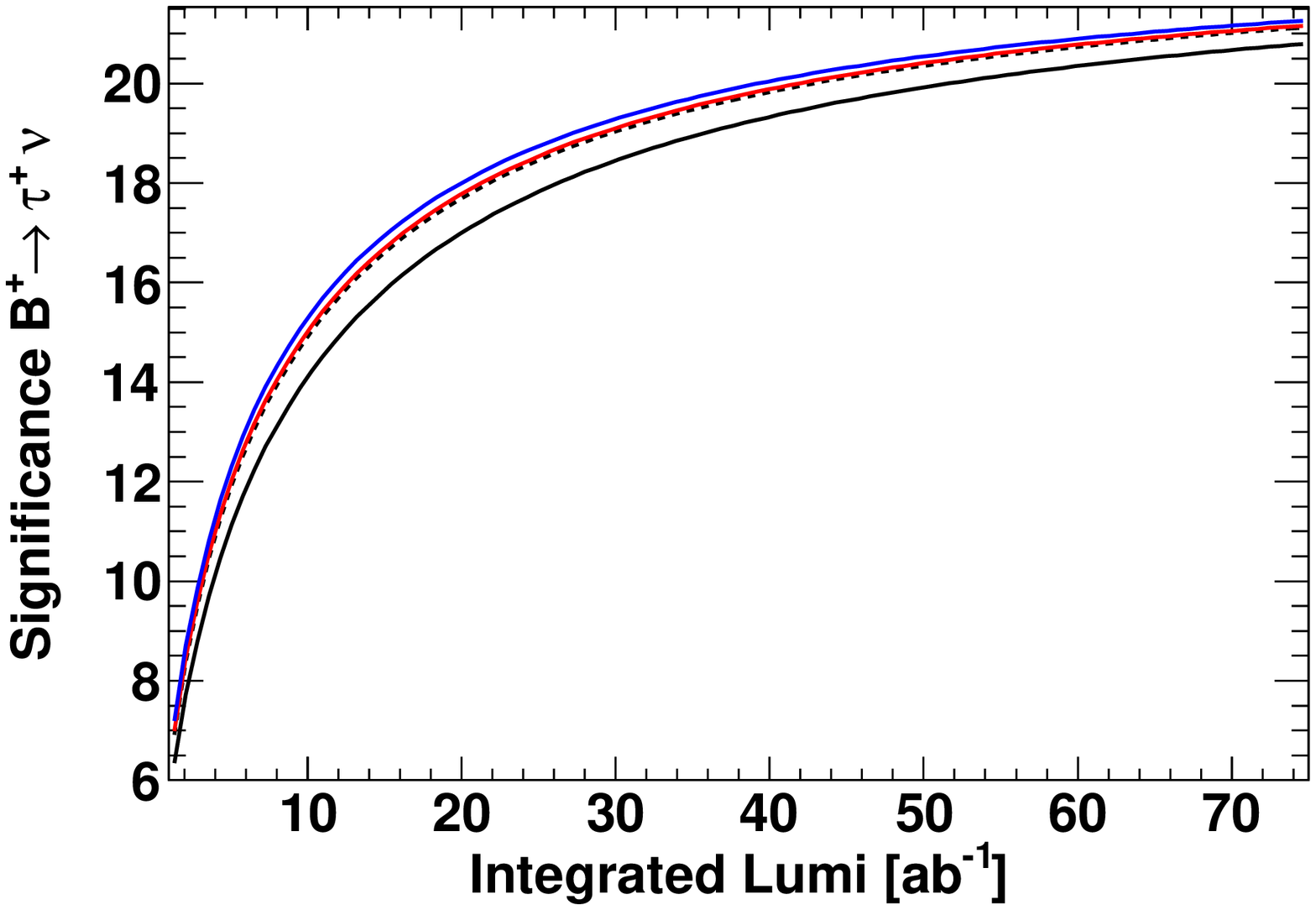}
\includegraphics[width=7.0cm,keepaspectratio]{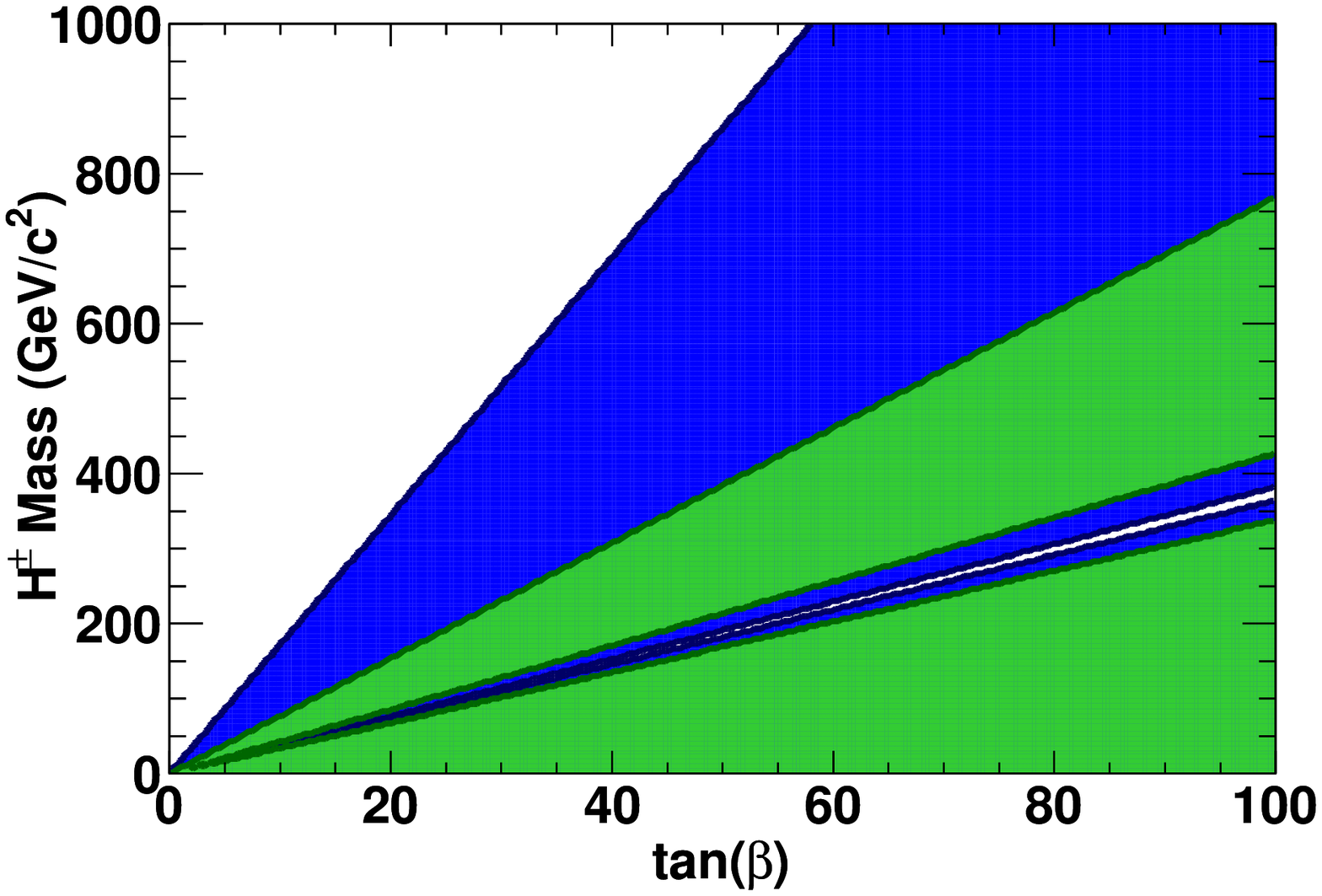}
\includegraphics[width=7.0cm,keepaspectratio]{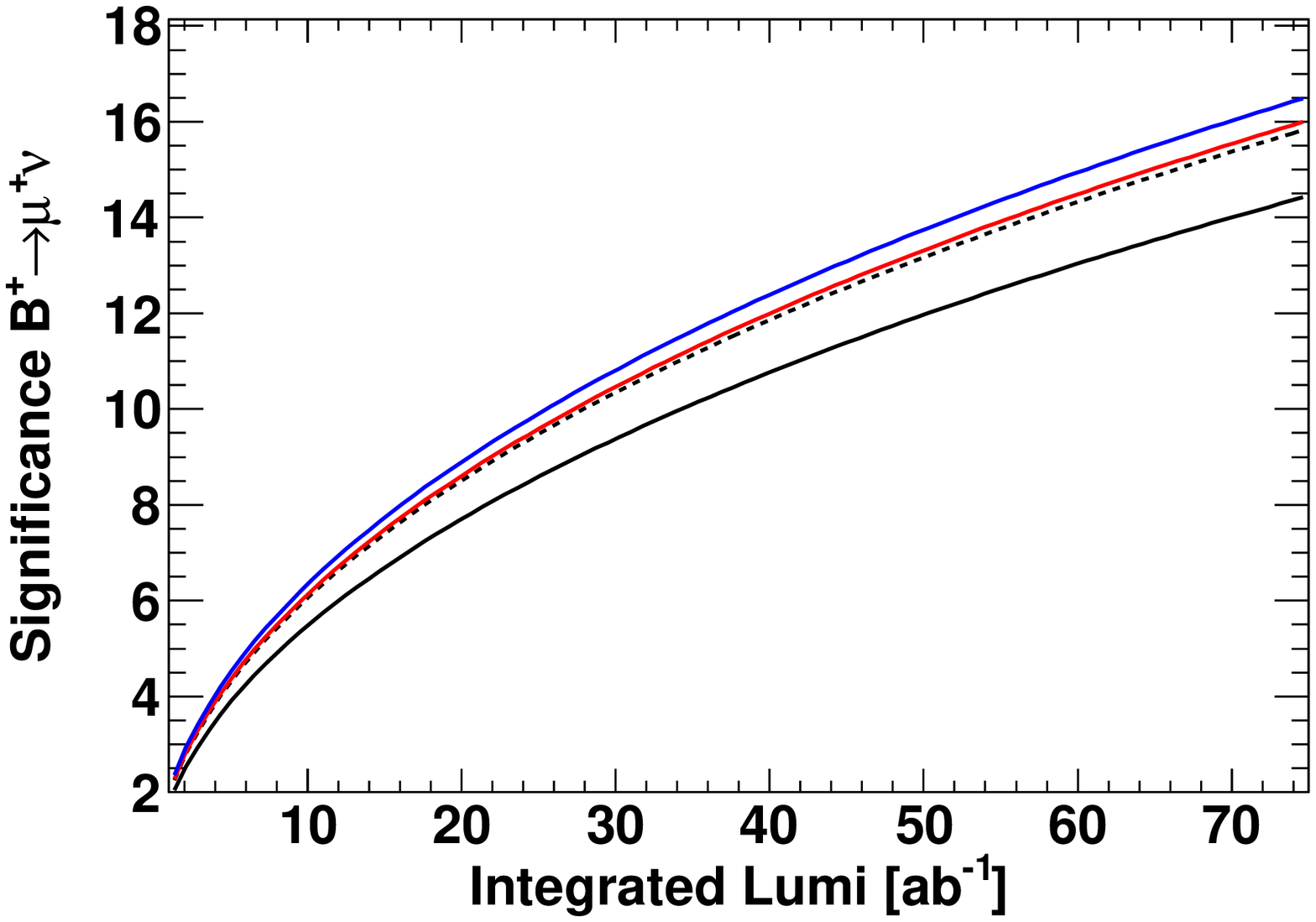}
\includegraphics[width=7.0cm,keepaspectratio]{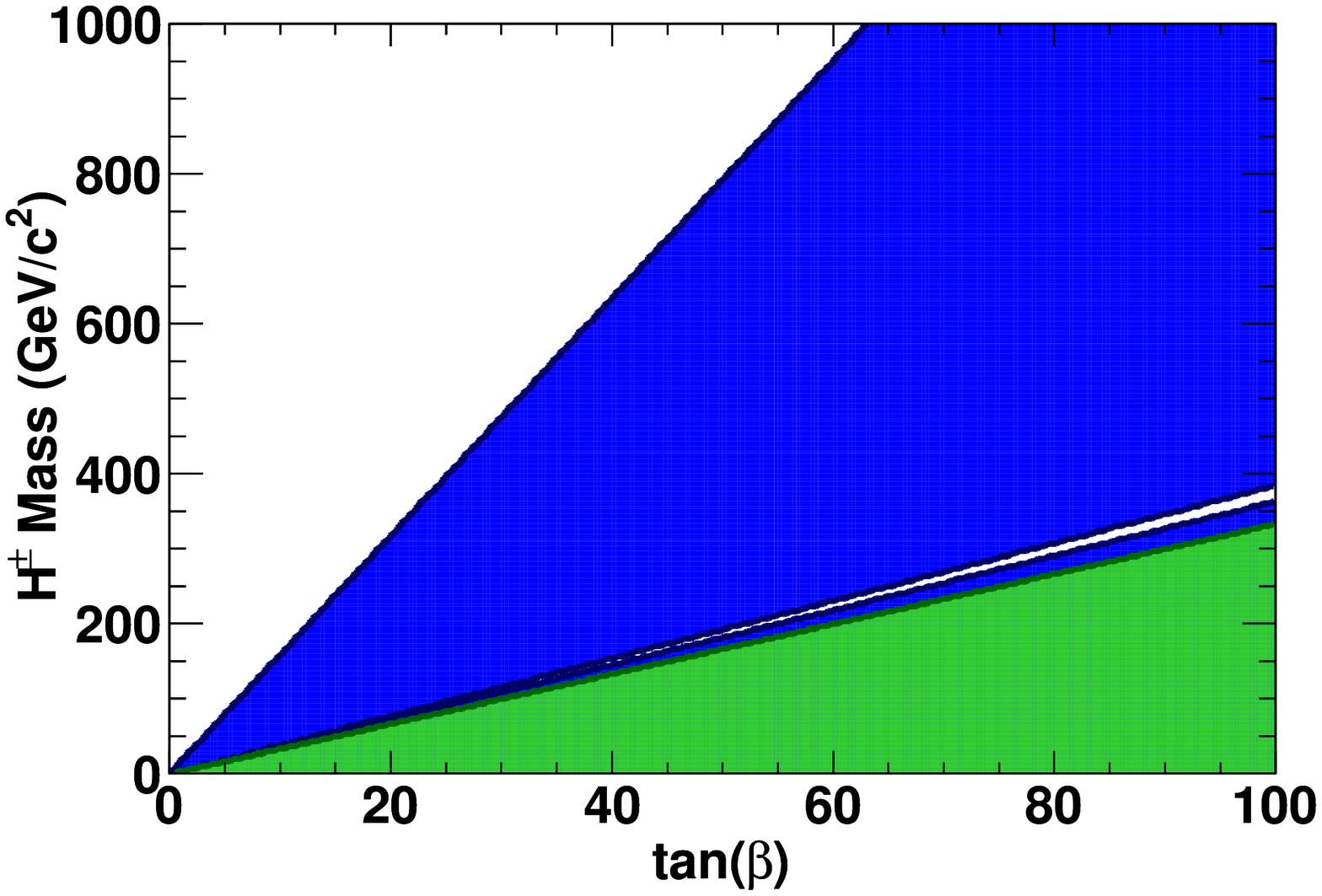}
\end{center}
\caption
{\label{fig:Ext_BToTauNu}
{\em left: Significance of the ${\rm BR}(B^+\rightarrow \tau^+\nu_{\tau})$ (top) and ${\rm BR}(B^+\rightarrow \mu^+\nu_{\mu})$ 
measurements as a function of integrated luminosity for the studied detector setups: BaBar (solid-black), 
SuperB baseline (dotted-black), Fwd-PID (red) and Bwd-EMC (blue). right: Excluded region on the $\tan\beta - M_H$ plane 
for the current (green) and expected sensitivities of SuperB at $75~{\rm ab}^{-1}$ (blue) from the 
${\rm BR}(B^+\rightarrow \tau^+\nu_{\tau})$ (top) and ${\rm BR}(B^+\rightarrow \mu^+\nu_{\mu})$ (bottom) measurements.
}}
\end{figure}

The irreducible systematic uncertainties on ${\rm BR}(B^+\rightarrow \tau^+\nu_{\tau})$ (mainly due to $B_{\rm tag}$ 
and $B_{\rm sig}$ reconstruction efficiencies and $B\bar{B}$ counting) is $8.7\%$, which is currently a factor of 
$\sim 2$ smaller than the statistical error. It is evident that this measurement will be systematic dominated in 
the near future if no effort is made to reduce the systematic error. The uncertainty will saturate at $\sim 9\%$ 
already at $\sim50~{\rm ab}^{-1}$, which is only $2/3$ of the total expected dataset of SuperB. Experience has 
shown that systematics can be reduced with higher statistics, as it is possible to study larger control samples. 
It is then assumed that the systematic uncertainty can be reduced by a factor of two, which can be considered as 
a moderately conservative scenario. Under this hypothesis, we obtain the top-left plot of 
figure~\ref{fig:Ext_BToTauNu}, where we show the statistical significance as a function of the integrated 
luminosity, which gives an uncertainty of $~4.5\%$ at $75~{\rm ab}^{-1}$. In order to translate this into an excluded 
region in the $\tan\beta - m_H$ plane, it is needed as well to make some hypothesis on the systematic error of the 
SM branching ratio (see eq.~\ref{eq:BF_Blnu_NP}). The main uncertainties coming from $|V_{ub}|$ and $f_B$ 
(see eq.~\ref{eq:BF_Blnu}), it will be assumed that the statistical error on $|V_{ub}|$ scales with luminosity and 
that the systematic component can be reduced by a factor of two. For the uncertainty on $f_B$ we use $1.5\%$, 
which is the expectated error for the SuperB era estimated by FLAG~\cite{FLAG}. In the right-top plot of figure~\ref{fig:Ext_BToTauNu} 
the excluded region (blue) for the expected sensitivities on the ${\rm BR}(B^+\rightarrow \tau^+\nu_{\tau})$ at 
$75~{\rm ab}^{-1}$ is shown. For comparison we show (green) the excluded region with the current uncertainties. As can be seen, 
the excluded region can be significantly increased with the expected sensitivities at SuperB full dataset.

In the case of the $B^+\rightarrow \mu^+\nu_{\mu}$ channel, the irreducible systematic uncertainty is $\sim4.0\%$. 
As with $B^+\rightarrow \tau^+\nu_{\tau}$ decay, it is assumed that the systemtic errors can be reduced by a factor of two 
for the SuperB era, which gives the left-bottom plot of figure~\ref{fig:Ext_BToTauNu}. As can be seen, the 
${\rm BR}(B^+\rightarrow \mu^+\nu_{\mu})$ measurement will not be systematic dominated in contrast to the 
${\rm BR}(B^+\rightarrow \tau^+\nu_{\tau})$. As shown in the bottom-right plot of figure~\ref{fig:Ext_BToTauNu}, 
the corresponding constraint on the $\tan\beta - M_H$ plane will be competitive with the one obtained from 
$B^+\rightarrow \tau^+\nu_{\tau}$ decays.

\section{Summary}

In summary, we have investigated the reach of SuperB in the search of the $B^+\rightarrow \ell^+\nu_{\ell}$ 
decays with both the hadronic and semi-leptonic techniques. Preliminary results based on the SuperB fast simulation 
have shown a significant increase on the signal to background ratio due to the boost reduction and the impact of the 
Fwd-PID and Bwd-EMC devices. It has also been shown that under moderately conservative hypothesis on the evolution of the 
systematic uncertainties both $B^+ \rightarrow \tau^+\nu_{\tau}$ and $B^+ \rightarrow \mu^+\nu_{\mu}$ decays will 
give competitive an unprecedent reduction of the NP parameter space ($\tan\beta - M_H$ plane) for the expected SuperB 
sensitivities at $75~{\rm ab^{-1}}$ of data.


\begin{thebibliography}{99}

\bibitem{SuperB} M.~Bone {\it et al.} [SuperB Collaboration], arXiv:0709.0451 [hep-ex].


\bibitem{Silverman} D. Silverman and H. Yao, Phys. Rev. D{\bf 38}, 214 (1988).
\bibitem{CKMmatrix} N. Cabibbo, Phys. Rev. Lett. {\bf 10}, 531 (1963); M. Kobayashind {\it et al.}, Prog. Theor. Phys. {\bf 49}, 652 (1973).
\bibitem{TauB} C. Amsler {\it et al.}, Physics Letters B {\bf 667}, 1 (2008).
\bibitem{HFAG} E. Barberio {\it et al.}, [Heavy Flavor Averaging Group], arXiv: hep-exp/0603003.
\bibitem{HPQCD} E. Gamiz, {\it et al.} [HPQCD Collaboration], Phys. Rev. D{\bf 80}, 014503 (2009).

\bibitem{TwoHiggsDoubletModel} W. S. Hou, Phys. Rev. D {\bf 48}, 2342 (1993).
\bibitem{SUSY} A. G. Akeroyd {\it et al.}, J. Phys G {\bf 29} 2311, (2003).

\bibitem{CKMFitter} J. Charles {\it et al.} [CKMfitter Group], Eur. Phys. J. C{\bf 41}, 1-131 (2005), (see http://ckmfitter.in2p3.fr).
\bibitem{UTFIT} M. Bona {\it et al.} [UTfit Collaboration], JHEP 0507 (2005) 028, (see http://www.utfit.org).

\bibitem{BaBar_BToenu_BTomunu_BTotaunu_SL} B. Aubert, {\it et al.} [BaBar Collaboration], Phys. Rev. D{\bf 81}:051101, 2010.
\bibitem{Belle_BTotaunu_SL} K. Hara, {\it et al.} [Bella Collaboration], arXiv: 1006.4201 [hep-ex].

\bibitem{BaBar_BTotaunu_HD} P. del Amo Sanchez, {\it et al.} [BaBar Collaboration], arXiv:1008.0104 [hep-ex].
\bibitem{Belle_BTotaunu_HD} K. Ikado, {\it te al.} [Bella Collaboration], Phys. Rev. Lett. {\bf 97}:251802, 2006.

\bibitem{BaBar_BToenu_BTomunu_HD} B. Aubert, {\it et al.} [BaBar Collaboration], Phys. Rev. D{\bf 79}:091101, 2009.
\bibitem{Belle_BToenu_BTomunu_HD} Phys. Lett. B {\bf 646}, 67 (2007).

\bibitem{FLAG} V. Lubicz, {\it "The CKM analysis: inputs from theory"}, talk given at {\it The Xth Nicola Cabibbo International 
Conference on Heavy Quarks and Leptons, October 11-15, 2010, Frascati - Italy}.

\end{thebibliography}
\end{document}